\newcommand{\version}{July 1, 2010}
\documentclass[a4paper,twoside,11pt]{article}
\usepackage[bindingoffset=0.45cm,textheight=22.5cm,hdivide={2.75cm,*,2.7cm}, vdivide={*,22cm,*}]{geometry}
\usepackage{amsmath}
\usepackage{amsfonts}
\usepackage{dsfont}
\let\mathbb=\mathds	

\usepackage[dvips,bookmarksnumbered=true,breaklinks=true]{hyperref}

\usepackage[numbers,square,comma,sort&compress]{natbib}

\newcommand{\uim}{UV/IR mixing}
\newcommand{\nc}{non-com\-mu\-ta\-tive}

\newcommand{\eqnref}[1]{Eqn.~(\ref{#1})}		
\newcommand{\secref}[1]{Section~\ref{#1}}		
\newcommand{\co}[2]{\left[#1,#2\right]}					
\newcommand{\aco}[2]{\left\{#1,#2\right\}}				
\newcommand{\starco}[2]{\left[ #1\stackrel{\star}{,}#2\right] }		
\newcommand{\var}[2]{\frac{\d #1}{\d #2}}				
\newcommand{\pa}{\partial}						
\newcommand{\ri}{\mathrm{i}}						
\renewcommand{\k}{\tilde{k}}						
\newcommand{\p}{\tilde{p}}						
\newcommand{\bc}{\bar{c}}						
\newcommand{\Act}{S}
\renewcommand{\a}{\alpha}
\renewcommand{\b}{\beta}
\newcommand{\g}{\gamma}
\renewcommand{\d}{\delta}

\renewcommand{\th}{\theta}
\newcommand{\eth}{\theta} 
\newcommand{\sth}{\varepsilon} 
\newcommand{\st}{\bar{\sigma}}

\newcommand{\m}{\mu}
\newcommand{\n}{\nu}

\renewcommand{\r}{\rho}
\newcommand{\s}{\sigma}

\newcommand{\w}{\omega}

\newcommand{\G}{\Gamma}

\renewcommand{\L}{\Lambda}

\newcommand{\W}{\Omega}

\newcommand{\inv}[1]{\frac{1}{#1}}				
\newcommand{\tinv}[1]{\tfrac{1}{#1}}
\newcommand{\intx}{\int\!\! {\rm d}^4x}						
\newcommand{\wsq}{\widetilde{\square}}
\newcommand{\ig}{\mathrm{i}g}

\newcommand{\bpsi}{\bar{\psi}}
\newcommand{\bB}{\bar{B}}
\newcommand{\bQ}{\bar{Q}}
\newcommand{\bJ}{\bar{J}}

\newcommand{\Tr}{\text{Tr}}
\newcommand{\id}{\mathds{1}}						
\newcommand{\R}{\mathds{R}}

\title{\begin{flushright}
       \small{UWThPh-2010-9}
       \end{flushright}
\vspace{3em}
A New Approach to Non-Commutative $U_\star(N)$ Gauge Fields
}
\author{Daniel N. Blaschke}

\date{\version}
\begin{document}
\maketitle
\thispagestyle{empty}
\begin{center}
\renewcommand{\thefootnote}{\fnsymbol{footnote}}
Faculty of Physics, University of Vienna\\
Boltzmanngasse 5, A-1090 Vienna (Austria)\\[0.3cm]
\ttfamily{E-mail: daniel.blaschke@univie.ac.at}
\vspace{0.5cm}
\end{center}%
\begin{abstract}
Based on the recently introduced model of Ref.~\cite{Blaschke:2009e} for {\nc} $U_\star(1)$ gauge fields, a generalized version of that action for $U_\star(N)$ gauge fields is put forward. In this approach to {\nc} gauge field theories, {\uim} effects are circumvented by introducing additional ``soft breaking'' terms in the action which implement an IR damping mechanism. The techniques used are similar to those of the well-known Gribov-Zwanziger approach to QCD.
\end{abstract}

%
%
\section{Introduction}
For a long time quantum field theories formulated in a Groenewold-Moyal deformed (or $\th$-deformed) space~\cite{Groenewold:1946,Moyal:1949} suffered from new types of divergences arising due to a phenomenon referred to as {\uim}~\cite{Minwalla:1999,Susskind:2000}. For a review on the topic see Refs.~\cite{Szabo:2001,Rivasseau:2007a,Blaschke:2010kw}. Only some years ago, Grosse and Wulkenhaar were able to resolve the {\uim} problem in the case of a scalar field theory by adding an oscillator-like term to the (Euclidean) action~\cite{Grosse:2003, Grosse:2004b}, thereby rendering it renormalizable to all orders of perturbation theory~\cite{Grosse:2004a,Rivasseau:2006a,Rivasseau:2006b}. Eventually, an alternative approach was put forward by Gurau et al.~\cite{Rivasseau:2008a} by replacing the oscillator term with one of type $\phi(-p)\inv{p^2}\phi(p)$. The authors were able to prove renormalizability of this ``$\inv{p^2}$-model'' to all orders by means of Multiscale Analysis. 

Inspired by these successes, similar approaches were tried for $U_\star(1)$ gauge theories\footnote{Note, that the non-commutativity of the space coordinates alters the gauge group, which is why I have denoted the deformed $U(1)$ group by $U_\star(1)$.} in Euclidean space~\cite{Grosse:2007,Wulkenhaar:2007,Blaschke:2007b,Wallet:2008a,Blaschke:2008a,Vilar:2009,Blaschke:2009b,Blaschke:2009e}. The latest approach, the model presented in Ref.~\cite{Blaschke:2009e}, seems to be a very promising candidate for a renormalizable $U_\star(1)$ gauge theory on $\th$-deformed space. 

In the present work I generalize this model to the $U_\star(N)$ gauge group. It is formulated on Euclidean $\R_\eth^4$ with the Moyal-deformed product 
\begin{align}
\starco{x_\m}{x_\n} \equiv  x_{\mu} \star
x_{\nu} -x_{\nu} \star x_{\mu} = \ri \sth\eth_{\mu \nu}\,,
\end{align}
of regular commuting coordinates $x_\mu$. The real parameter $\sth$ has mass dimension $-2$, rendering the constant antisymmetric matrix $\eth_{\m\n}$ dimensionless.

In the following I will use the abbreviations $\tilde{v}_\m \equiv \eth_{\m\n}v_\n$ for vectors $v$ and $\tilde{M} \equiv \eth_{\m\n}M_{\m\n}$ for matrices $M$. For the deformation, I furthermore consider the simplest block-diagonal form
\begin{align}\label{eq:def-eth}
\eth_{\mu\nu}=\left(\begin{array}{cccc}
0&1&0&0\\
-1&0&0&0\\
0&0&0&1\\
0&0&-1&0
\end{array}\right)\, ,
\end{align}
for the dimensionless matrix describing non-commutativity.

\section{The \texorpdfstring{$U_\star(N)$}{U(N)} gauge field action}
The main ideas that in a series of papers~\cite{Blaschke:2008a,Blaschke:2009a,Vilar:2009,Blaschke:2009b,Blaschke:2009c,Blaschke:2009d} led to the construction of a model for $U_\star(1)$ gauge fields~\cite{Blaschke:2009e}, which has a good chance of being fully renormalizable, are:
\begin{itemize}
\item to implement a damping mechanism similar to the one present in the scalar $\inv{p^2}$-model of Gurau et al.~\cite{Rivasseau:2008a},
\item to endow the tree level action with counter terms for the quadratic and linear one-loop infrared divergent terms of type~\cite{Hayakawa:1999,Armoni:2000xr,Ruiz:2000,Blaschke:2005b}:
\begin{align}\label{eq:generic-IR-div}
\Pi^{\text{IR}}_{\m\n}(k)&\propto\frac{\k_\m\k_\n}{(\sth\k^2)^2}\,,
\end{align}
and
\begin{align}
\Gamma^{3A,\text{IR}}_{\m\n\r}(p_1,p_2,p_3)
&\propto\cos\left(\sth \frac{p_1\p_2}{2}\right)\sum\limits_{i=1,2,3}\frac{\p_{i,\m}\p_{i,\n}\p_{i,\r}}{\sth(\p_i^2)^2}\,,
\label{eq:counterterm-3A}
\end{align}
\item and to keep the model as simple as possible.
\end{itemize}
The greatest difficulty in the early approaches turned out to be the implementation of the IR damping: In Ref.~\cite{Blaschke:2008a} an additional gauge invariant term was added to the action which ultimately led to an infinite number of vertices. Localization of that term through the introduction of auxiliary fields\footnote{By ``localization'' it is only meant that the inverse of covariant derive operators leading to infinitely many vertices no longer enters the action explicitly. Of course the star products remain non-local nonetheless.} could remedy the situation with respect to the tree level vertices. However, other problems concerning renormalizability due to additional new Feynman rules for the auxiliary fields, remained --- cf.~\cite{Vilar:2009,Blaschke:2009b,Blaschke:2009d}. Therefore, an alternative approach was proposed in Ref.~\cite{Blaschke:2009e}, where the required extension to the $U_\star(1)$ gauge field action was implemented by means of a ``soft breaking'' technique similar to the Gribov-Zwanziger action in QCD --- see Refs.~\cite{Gribov:1978,Zwanziger:1989,Zwanziger:1993,Baulieu:2009} for details. 

Here I will follow the same ideas. Furthermore, it must also be taken into account, that in {\nc} $U_\star(N)$ gauge field models only the $U_\star(1)$ subsector is responsible for {\uim} (cf. Refs.~\cite{Armoni:2000xr,Armoni:2001,Armoni:2002fh}). This means, that infrared divergent terms only appear in Feynman graphs which have at least one external leg in the $U_\star(1)$ subsector. The key point here is that, by employing the soft breaking mechanism, one only modifies the infrared regime of the model while keeping the UV intact. Both UV divergences, on the one hand, as well as IR terms originating from {\uim} in e.g. one-loop corrections, on the other hand, are caused by the UV regime of the integrand in a Feynman loop graph~\cite{Minwalla:1999,Susskind:2000,Szabo:2001}. Therefore, the same one-loop results to leading order are expected for the current model as in the literature~\cite{Armoni:2000xr} (see the discussion in \secref{sec:one-loop} below and in Ref.~\cite{Blaschke:2009e}). In other words, we need to implement counter terms of type \eqref{eq:generic-IR-div} and \eqref{eq:counterterm-3A} in the soft breaking part of our action for $U_\star(1)$ gauge fields, but not for the pure $SU_\star(N)$ sector.

\paragraph{Notation.}
Throughout the remainder of this paper, the following notation will be used: Following Ref.~\cite{Armoni:2000xr} I denote $U_\star(N)$ indices with capital letters $A,B,C,\ldots$ and $SU_\star(N)$ indices with $a,b,c,\ldots$. Finally, the index $0$ is used for fields which are $U_\star(1)$, and whenever an index is omitted, the according field including the $U(N)$ gauge group generator $T^A$ is meant. Furthermore, all products are implicitly assumed to be deformed (i.e. star products).

\paragraph{$\mathbf{U_\star(N)}$ gauge fields.}
The covariant derivative $D_\m$ and the field strength $F_{\m\n}$ are defined as
\begin{align}
D_\m\bullet&=\pa_\m\bullet-\ig\co{A_\m}{\bullet}\,, & A_\m&=A_\m^AT^A\,, \nonumber\\
F_{\m\n}&=\pa_\m A_\n-\pa_\n A_\m-\ig\co{A_\m}{A_\n}\,,
\end{align}
where $T^A$ are the generators of the $U(N)$ gauge group. They are normalized as $\Tr(T^AT^B)=\inv{2}\d^{AB}$, and $T^0=\inv{\sqrt{2N}}\id_N$ (cf.~\cite{Armoni:2000xr}). 
Due to the star product, the field strength tensor $F_{\m\n}$ exhibits additional couplings between the $U_\star(1)$ and the $SU_\star(N)$ sector, i.e. we have
\begin{align}
F_{\m\n}&=\left(\pa_\m A^0_\n-\pa_\n A^0_\m-\frac{\ig}{2}d^{AB0}\co{A^A_\m}{A^B_\n}\right)T^0\nonumber\\
&\quad +\left(\pa_\m A^c_\n-\pa_\n A^c_\m+\frac{g}{2}f^{abc}\aco{A^a_\m}{A^b_\n}-\frac{\ig}{2}d^{ABc}\co{A^A_\m}{A^B_\n}\right)T^c\nonumber\\
&\equiv F^0_{\m\n}T^0+F^c_{\m\n}T^c \,,
\end{align}
where $f^{abc}$ and $d^{ABC}$ are (anti)symmetric structure constants of the gauge group. 
The terms proportional to $d^{AB0}=\sqrt{\frac{2}{N}}\d^{AB}$ contain both types of fields, i.e. $U_\star(1)$ and $SU_\star(N)$, and hence giving rise to the additional couplings. 
In the commutative limit, the star commutators would vanish and the two sectors would decouple once more.

Similarly, one has for the covariant derivative of e.g. a ghost field $c$:
\begin{align}
D_\m c&=\left(\pa_\m c^0-\frac{\ig}{2}d^{AB0}\co{A^A_\m}{c^B}\right)T^0 +\left(\pa_\m c^c+\frac{g}{2}f^{abc}\aco{A^a_\m}{c^b}-\frac{\ig}{2}d^{ABc}\co{A^A_\m}{c^B}\right)T^c
\,.
\end{align}

\paragraph{Proposed action.}
In light of the considerations above, the following generalized BRST invariant $U_\star(N)$ gauge field action formulated in Euclidean $\R_\eth^4$ including additional $U_\star(1)$ auxiliary fields is suggested:
{\allowdisplaybreaks
\begin{align}\label{eq:renormalizable_action}
\Act&=\Act_{\text{inv}}+\Act_{\text{gf}}+\Act_{\text{aux}}+\Act_{\text{soft}}+\Act_{\text{ext}}\,,\nonumber\\*
\Act_{\text{inv}}&=\intx\tinv{4}F^A_{\m\n}F^A_{\m\n}\,,\nonumber\\*
\Act_{\text{gf}}&=\intx\,s\left(\bc^A\,\pa_\m A^A_\m\right)=\intx\left(b^A\,\pa_\m A^A_\m-\bc^A\,\pa_\m (D_\m c)^A\right)\,,\nonumber\\
\Act_{\text{aux}}&=-\intx\,s\left(\bpsi^0_{\m\n}B^0_{\m\n}\right)=\intx\left(-\bB^0_{\m\n}B^0_{\m\n}+\bpsi^0_{\m\n}\psi^0_{\m\n}\right)\,,\nonumber\\
\Act_{\text{soft}}&=\intx\,s\Bigg[\!\left(\bar{Q}^0_{\m\n\a\b}B^0_{\m\n}+Q^0_{\m\n\a\b}\bB^0_{\m\n}\right)\inv{\wsq}\left(\!f^0_{\a\b}+\s\frac{\eth_{\a\b}}{2}\tilde{f}^0\!\right)\nonumber\\*
&\quad\qquad\qquad
+e^{ABC}Q'^0\aco{A^A_\m}{A^B_\n} \frac{\tilde{\pa}_\m\tilde{\pa}_\n\tilde{\pa}_\r}{\wsq^2}A^C_\r\Bigg]\nonumber\\*
&=\intx\bigg[\!\!\left(\bar{J}^0_{\m\n\a\b}B^0_{\m\n}+J^0_{\m\n\a\b}\bB^0_{\m\n}\right)\!\inv{\wsq}\!\left(\!f^0_{\a\b}+\s\frac{\eth_{\a\b}}{2}\tilde{f}^0\!\right)\! - \bar{Q}^0_{\m\n\a\b}\psi^0_{\m\n}\inv{\wsq}\!\left(\!f^0_{\a\b}+\s\frac{\eth_{\a\b}}{2}\tilde{f}^0\!\right)\nonumber\\*
& \qquad \qquad -\left(\bar{Q}^0_{\m\n\a\b}B^0_{\m\n}+Q^0_{\m\n\a\b}\bar{B}^0_{\m\n}\right)\inv{\wsq}\mathop{s}\left(\!f^0_{\a\b}+\s\frac{\eth_{\a\b}}{2}\tilde{f}^0\!\right)\nonumber\\*
&\qquad\qquad
+e^{ABC}J'^0\aco{A^A_\m}{A^B_\n} \frac{\tilde{\pa}_\m\tilde{\pa}_\n\tilde{\pa}_\r}{\wsq^2}A^C_\r -
e^{ABC}Q'^0s\left(\aco{A^A_\m}{A^B_\n} \frac{\tilde{\pa}_\m\tilde{\pa}_\n\tilde{\pa}_\r}{\wsq^2}A^C_\r\right)\!\bigg]\,, \nonumber\\
\Act_{\text{ext}}&=\intx\left(\W^A_\m (sA_\m)^A+\w^A (sc)^A\right)\,,
\end{align}
where 
}
\begin{align}\label{eq:def-eABC}
e^{ABC}&\equiv d^{ABC}-d^{abc}\d^{aA}\d^{bB}\d^{cC}\,,
&& \wsq = \tilde\partial_\mu \tilde\partial_\mu \,.
\end{align}
The abbreviation $e^{ABC}$ denotes all symmetric structure constants $d^{ABC}$ where at least one index is $0$, i.e. in the $U_\star(1)$ subsector of the gauge group. 
The reason for this restriction has already been mentioned above: In loop calculations terms of type \eqref{eq:counterterm-3A} only appear when at least one external leg is in the $U_\star(1)$ subsector, as was first shown by A. Armoni~\cite{Armoni:2000xr,Armoni:2001}.

Finally, $f^0_{\m\n}$ denotes the free part of $F^0_{\m\n}$, i.e.
\begin{align}
f^0_{\m\n}&=\pa_\m A^0_\n-\pa_\n A^0_\m
\,, &
\tilde{f}^0&=\eth_{\m\n}f^0_{\m\n}\,,
\end{align}
the multiplier field $b$ implements the Landau gauge fixing $\pa_\m A_\m=0$, $\bc$/$c$ denote the (anti)ghost, and $\s$ is a dimensionless parameter. The complex $U_\star(1)$ field $B^0_{\m\n}$, its complex conjugate $\bB^0_{\m\n}$ and the associated additional ghosts $\bpsi^0$, $\psi^0$ are introduced in order to implement the IR damping mechanism explained in Ref.~\cite{Blaschke:2009e} on the according $U_\star(1)$ gauge model. The additional $U_\star(1)$ sources $\bQ^0,Q^0,Q'^0,\bJ^0,J^0,J'^0$ are needed in order to ensure BRST invariance of the action in the ultraviolet. In the infrared they take the ``physical'' values
\begin{align}
&\bQ^0_{\m\n\a\b}\Big|_{\text{phys}}=Q^0_{\m\n\a\b}\Big|_{\text{phys}}=Q'^0\Big|_{\text{phys}}=0\,,  
&& J'^0\Big|_{\text{phys}}=\ri g\g'^2\,,\nonumber\\
&\bJ^0_{\m\n\a\b}\Big|_{\text{phys}}=J^0_{\m\n\a\b}\Big|_{\text{phys}}=\frac{\g^2}{4}\left(\d_{\m\a}\d_{\n\b}-\d_{\m\b}\d_{\n\a}\right)\,,
\label{eq:physical-values}
\end{align}
where $\g$ and $\g'$ are Gribov-like parameters of mass dimension 1 (cf.~\cite{Gribov:1978,Zwanziger:1989,Zwanziger:1993,Baulieu:2009}). 
The action \eqref{eq:renormalizable_action} is hence invariant under the BRST transformations
\begin{align}\label{eq:BRST_of_renorm_action}
&sA_\mu=D_\mu c\,,  &&  sc=\ig {c}{c}\, ,\nonumber\\
&s\bc=b\,,          && sb=0\, ,  \nonumber\\
&s\bpsi_{\mu\nu}=\bB_{\m\n}\,,     && s\bB_{\m\n}=0\,,\nonumber\\
&sB_{\m\n}=\psi_{\m\n}\,,     && s\psi_{\m\n}=0\,,\nonumber\\
&s\bar{Q}_{\m\n\a\b}=\bar{J}_{\m\n\a\b}\,, && s\bar{J}_{\m\n\a\b}=0\,, \nonumber\\
&sQ_{\m\n\a\b}=J_{\m\n\a\b}\,, && sJ_{\m\n\a\b}=0\,,\nonumber\\
&sQ'=J'\,, && sJ'=0\,,
\end{align}
and for the non-linear transformations $sA_\m$ and $sc$, external sources $\W_\m$ and $\w$ have been introduced, respectively. 
Notice, that the auxiliary fields form BRST doublets reflecting their unphysical nature. Dimensions and ghost numbers of the fields involved are given in Table~\ref{tab:field_prop}.
\begin{table}[ht]
\caption{Properties of fields and sources.}
\label{tab:field_prop}
\centering
\begin{tabular}{lcccccccccccccccc}
\hline
\hline
\rule[12pt]{0pt}{0.1pt}
Field       & $A$ & $c$ & $\bc$ & $B$ & $\bB$ & $\psi$ & $\bpsi$ & $J$ & $\bar{J}$ & $J'$ & $Q$ & $\bar{Q}$ & $Q'$ & $\W$ & $\w$ & b \\[2pt]
\hline
$g_\sharp$  &    0   &  1  &   -1  &     0      &    0         &        1      &      -1        &   0            &       0      &  0        &  -1     &   -1       &  -1 &   -1 &  -2  & 0 \\
Mass dim.   &    1   &  0  &   2   &     2      &    2         &        2      &       2        &   2     &  2  &  2       &       2              &  2             &  2 & 3 & 4 & 2\\
\hline
\hline
\end{tabular}
\end{table}
Finally, the Slavnov-Taylor identity describing the BRST symmetry content of the model is given by
\begin{align}
\mathcal{B}(\Act)&=\intx\Bigg(\var{\Act}{\W_\m}\var{\Act}{A_\m}+\var{\Act}{\w}\var{\Act}{c}+b\var{\Act}{\bc}+\bB_{\m\n}\var{\Act}{\bpsi_{\m\n}}+\psi_{\m\n}\var{\Act}{B_{\m\n}}\nonumber\\*
&\phantom{=\intx\Bigg(} +\bar{J}_{\m\n\a\b}\var{\Act}{\bar{Q}_{\m\n\a\b}}+J_{\m\n\a\b}\var{\Act}{Q_{\m\n\a\b}}+J'\var{\Act}{Q'}\Bigg)=0\,.
\end{align}

\section{Discussion of one-loop properties}\label{sec:one-loop}
The gauge field propagator (using \eqnref{eq:physical-values}, cp. Ref.~\cite{Blaschke:2009e}) takes the form
\begin{align}\label{eq:prop_aa}
 G^{A^0A^0}_{\m\n}(k)&=\left[k^2+\frac{\g^4}{\k^2}\right]^{-1}\left[\d_{\m\n}-\frac{k_\m k_\n}{k^2}-\frac{\st^4}{\left(k^2+\left(\st^4+\g^4\right)\inv{\k^2}\right)}\frac{\k_\m\k_\n}{(\k^2)^2}\right]
\,, \nonumber\\
  G^{A^aA^b}_{\m\n}(k)&=\frac{\d^{ab}}{k^2}\left(\d_{\m\n}-\frac{k_\m k_\n}{k^2}\right)\,,
\end{align}
where we have introduced the abbreviation
\begin{align}
\st^4\equiv 2(1+\s)\s\g^4\,,
\end{align}
and considered the case where $\eth_{\m\n}$ has the simple block diagonal form given in \eqref{eq:def-eth} so that $\k^2=k^2$ and $\th_{\m\n}\th_{\m\n}=4$. Notice that the soft-breaking terms in the action \eqref{eq:renormalizable_action} lead to an IR modified propagator in the $U_\star(1)$ sector. Two limits are of special interest: the IR limit $k^2\to0$ and the UV limit $k^2\to\infty$. A simple analysis reveals that
\begin{equation}
G^{A^0A^0}_{\m\n}(k)\approx\begin{cases}\frac{\k^2}{\g^4}\left[\d_{\m\n}-\frac{k_\m k_\n}{k^2}-\frac{\st^4}{\left(\st^4+\g^4\right)}\frac{\k_\m\k_\n}{\k^2}\right], & \text{for } \k^2\to0\,,\\[0.8em]
 \inv{k^2}\left(\d_{\m\n}-\frac{k_\m k_\n}{k^2}\right), & \text{for } k^2\to\infty\,.\\
\end{cases}
\label{eq:prop_aa_limits}
\end{equation}
From \eqnref{eq:prop_aa_limits} one can nicely see the appearance of a term of the same type as \eqref{eq:generic-IR-div} in the IR limit. This, by construction, admits the absorption of the problematic divergent terms appearing in the one loop results~\cite{Blaschke:2009d}. Another advantageous property of the gauge propagator is that the UV limit (from which divergences originate), admits to neglect the term proportional to $\g$ which reduces the number of terms in Feynman integrals considerably, especially since both $U_\star(1)$ and $SU_\star(N)$ propagators are of the same form in this limit.

The ghost propagator takes the usual simple form $G^{\bc c}(k)=-\frac{\d^{AB}}{k^2}$. 
Since $A_\m$ does not couple to the auxiliary fields ($B,\bB,\psi,\bpsi$) no other propagator will contribute to physical results, and they are hence omitted at this point.

Additionally, the model \eqref{eq:renormalizable_action} features several vertices. Similar to \eqref{eq:prop_aa_limits}, one may consider their UV approximations, which for the purpose of one-loop calculations would be sufficient. Within this approximation the vertices are given by the same expressions as in e.g.~\cite{Armoni:2000xr}. 
In other words, terms proportional to $\g'$ in the $3A$ vertex may be omitted as they are negligible for large momenta and hence play no role in the one-loop divergences. In the infrared, they scale as \eqref{eq:counterterm-3A}, but such linear IR divergent terms are always compensated by quadratically IR damping gauge field propagators --- cp. \eqnref{eq:prop_aa_limits}.

Considering the scaling behaviour of all these Feynman rules for large momenta, one derives an estimate for the superficial degree of ultraviolet divergences, which is the well-known result 
\begin{align}\label{eq:powercounting}
 d_\gamma=4-E_A-E_{c\bc}\,, 
\end{align}
where $E$ denotes the number of external legs of the various field types in a Feynman graph. 

When extracting the resulting UV divergences of the various one-loop corrections, only those terms in the Feynman rules contribute  which survive the UV approximations discussed above. Hence, these computations reduce to exactly the same ones already done in the literature, i.e. see~\cite{Armoni:2000xr,Ruiz:2000,Hayakawa:1999,Blaschke:2009e}. The vacuum polarization hence exhibits a logarithmic UV divergence of the form
\begin{align}
\Pi^{\text{UV}}_{\m\n}&\propto \d^{AB}g^2\left(p^2\d_{\m\n}-p_\m p_\n\right)\ln|\L^2\p^2|+\text{finite},
\end{align}
where $\L$ denotes an ultraviolet cutoff. In addition, as discussed in the literature, there is also a UV finite contribution to the vacuum polarization which diverges quadratically for vanishing external momentum. It is of type \eqref{eq:generic-IR-div} and appears \emph{only} in graphs where the external legs are in the $U_\star(1)$ sector. In fact, this can be easily seen by considering the according phase factors when the free colour indices $a,b\in SU_\star(N)$. In that case, one has phase factors of the form
\begin{align}
d^{aCD}d^{bCD}\sin^2(k\p/2)+f^{acd}f^{bcd}\cos^2(k\p/2)=N\d^{ab}\,,
\end{align}
since $d^{aCD}d^{bCD}=f^{acd}f^{bcd}=N\d^{ab}$. Clearly, they are phase-independent and hence lead to purely planar contributions. 

The $3A$-vertex corrections exhibit UV divergences of the form~\cite{Armoni:2000xr,Blaschke:2009e}
\begin{align}\label{3A_correction_UV}
\Gamma^{3A,\text{UV}}_{\m\n\r}(p_1,p_2,p_3)&\propto-g^2N\ln (\Lambda) \widetilde{V}^{3A,\text{tree}}_{\m\n\r}(p_1,p_2,p_3)\,,
\end{align}
as well as finite contributions. The latter exhibit IR divergences in the external momenta of type \eqref{eq:counterterm-3A} only if at least one external leg is in the $U_\star(1)$ subsector, as emphasized above~\cite{Armoni:2000xr,Armoni:2002fh}. 
In fact, one can argue that infrared divergent terms appear also in $n$-point graphs \emph{only} if at least one of the external legs is in the $U_\star(1)$ subsector by considering, for example, graphical representations of non-planar Feynman diagrams using the 't Hooft double index notation~\cite{Armoni:2001,Armoni:2002fh,Levell:2003ta}. What one finds, is that it is impossible to construct a non-planar graph without having at least one external $U_\star(1)$ leg, and since {\uim} occurs only in non-planar graphs~\cite{Minwalla:1999,Susskind:2000,Szabo:2001}, as is well-known, the same is true for the appearance of IR terms.

\section{Renormalization}
A renormalizable action must be form-invariant under quantum corrections, and its parameters are fixed by renormalization conditions on the vertex functions. So far, we have worked in Landau gauge with $\a=0$ (cf. \eqnref{eq:prop_aa}). However, for the following considerations, where we follow the steps of Ref.~\cite{Blaschke:2009e}, an arbitrary gauge parameter $\a\neq0$ will be more advantageous\footnote{Note, that the quadratic IR divergence is independent of the gauge fixing~\cite{Blaschke:2005b,Ruiz:2000,Hayakawa:1999b}.}, as the inverse of the gauge field propagator diverges in the limit $\a\to0$ due to elimitaion of the multiplier field $b$. 

\subsection{The renormalized propagator}
Recall that the tree-level gauge field propagator \eqref{eq:prop_aa} (using a compact and intuitive notation) has the form
\begin{align}\label{eq:propAA-general}
G^{A^AA^B}_{\m\n}(k)&=\frac{\d^{AB}}{k^2\mathcal{D}(k)}\left(\d_{\m\n}-\left(1-\a\mathcal{D}(k)\right)\frac{k_\m k_\n}{k^2}-\mathcal{F}(k)\frac{\k_\m\k_\n}{\k^2}\right)\,,
\end{align}
where we have introduced the abbreviations
\begin{align}
\mathcal{D}(k)&\equiv\left(1+\d^{A0}\d^{B0}\frac{\g^4}{(\k^2)^2}\right)
\,, &
\mathcal{F}(k)&\equiv\frac{\d^{A0}\d^{B0}}{\k^2}\frac{\st^4}{\left(k^2+\left(\st^4+\g^4\right)\inv{\k^2}\right)}\,,
\end{align}
i.e. terms including parameters $\st$ or $\g$ only appear in the $A^0A^0$-propagator. 
Its inverse, the tree-level two-point vertex function, is given by
\begin{align}
 \G^{AA,\text{tree}}_{\m\n}(k)&=\left(G_{AA}^{-1}\right)_{\m\n}(k)=\d^{AB}k^2\mathcal{D}(k)\left(\d_{\m\n}+\left(\inv{\a\mathcal{D}(k)}-1\right)\frac{k_\m k_\n}{k^2}+\frac{\d^{A0}\d^{B0}\st^4}{k^2\k^2\mathcal{D}(k)}\frac{\k_\m\k_\n}{\k^2}\right)\,.
\end{align}
As discussed in Section~\ref{sec:one-loop}, its (divergent) one-loop corrections are qualitatively given by
\begin{align}
 \G^{AA,\text{corr.}}_{\m\n}(k)&=\varPi_1\frac{\k_\m\k_\n}{(\k^2)^2}+\varPi_2\left(k^2\d_{\m\n}-k_\m k_\n\right)\,,\nonumber\\
\textrm{with }\quad\varPi_1&\propto\d^{A0}\d^{B0}\frac{Ng^2}{\sth^2}
\,,\qquad
\varPi_2\propto\d^{AB}{Ng^2}\ln\L\,,
\end{align}
where $\L$ is an ultraviolet cutoff. Hence, we find that in 
introducing the wave-function renormalization $Z_A$ and the renormalized parameters $\g_r$ and $\st_r$ according to
\begin{align}\label{eq:renormalized_parameters}
 Z_A&=\inv{\sqrt{1-\varPi_2}}
\,, &
\g_r^4&={\g^4}{Z_A^2}
\,, &
\st_r^4&=\left({\st^4-\varPi_1}\right){Z_A^2}\,,
\end{align}
the one-loop two-point vertex function can be cast into the same form as its tree-level counter part, i.e.
\begin{align}
 \G^{AA,\text{ren}}_{\m\n}(k)&=\G^{AA,\text{tree}}_{\m\n}(k)-\G^{AA,\text{corr.}}_{\m\n}(k)\nonumber\\
&=\frac{k^2\mathcal{D}_r}{Z_A^2}\left(\d_{\m\n}+\left(\frac{Z_A^2}{\a\mathcal{D}_r}-1\right)\frac{k_\m k_\n}{k^2}+\frac{\d^{A0}\d^{B0}\st_r^4}{k^2\k^2\mathcal{D}_r}\frac{\k_\m\k_\n}{\k^2}\right)\,,\nonumber\\
\mathcal{D}_r(k)&\equiv\left(1+\d^{A0}\d^{B0}\frac{\g_r^4}{(\k^2)^2}\right)\,.
\end{align}
Perhaps the most important result of this calculation is that the wave-function renormalization $Z_A$ is exactly the same for the $U_\star(1)$ and the $SU_\star(N)$ gauge field because it is independent of $\varPi_1$. 
In fact, the quadratic IR divergence $\varPi_1$ only enters the renormalization of the newly introduced parameter $\st$. 
For the sake of completeness, we note that the renormalized propagator takes the same form as \eqref{eq:propAA-general} apart from an additional prefactor $Z_A^2$, but with all parameters replaced by their renormalized counter parts.

We also need to provide renormalization conditions for the two-point vertex function for the gauge boson
\begin{align}
\Gamma^{AA}_{\mu\rho} 
& = \Gamma^{AA,T} (\d_{\m\r} - \frac{k_\m k_\r}{k^2}) +(\Gamma^{AA,NC})\, \frac{\tilde k_\m \tilde k_\r}{\tilde k^2}  + (\Gamma^{AA,L})\, \frac{k_\m k_\r}{k^2}\,,
\end{align}
where the vertex function has been split into a transversal and longitudinal part following Ref.~\cite{Blaschke:2009e}. We have used the identifications
\begin{align}
\Gamma^{AA,T} & = k^2 \mathcal D\,,\qquad
\Gamma^{AA,NC} =  \d^{A0}\d^{B0}\frac{\st^4}{\k^2}\,,\qquad (\Gamma^{AA,L}) = \frac{k^2}{\alpha} \,,
\end{align}
which finally allow to formulate the following renormalization conditions:
{\allowdisplaybreaks
\begin{align}
\frac{(\k^2)^2}{k^2} \Gamma^{AA,T} \Big|_{k^2=0} & = \d^{A0}\d^{B0}\gamma^4\,,
&
\inv{2 k^2} \frac{\pa (k^2 \Gamma^{AA,T})}{\pa k^2} \Big|_{k^2=0} & = 1
\,, \nonumber\\
\Gamma^{AA,L}\Big|_{k^2=0} & = 0\,,
&
\frac{\pa \Gamma^{AA,L}}{\pa k^2} \Big|_{k^2=0} & = \inv{\a}
\,, \nonumber\\
\k^2 \Gamma^{AA,NC}\Big|_{k^2=0} & = \d^{A0}\d^{B0}\st^4
\,.
\end{align}
}

\subsection{The \texorpdfstring{$\beta$}{beta}-function and renormalization of \texorpdfstring{$\g'$}{gamma-prime}}
As usual, the $\beta$-function is given by the logarithmic derivative of the bare coupling $g$ with respect to the cut-off for fixed $g_r$, i.e.
\begin{align}\label{eq:betafunction_def}
\b(g,\L) &=  \L \frac{ \pa g }{ \pa \L}\Big|_{g_r\, \text{fixed}}
\,, &
\b(g) &= \lim_{\L\to\infty} \b(g,\L)\,.
\end{align}
The renormalized coupling is obtained from the relation $g_r = g Z_g Z_A^3$, where $Z_g$ denotes the multiplicative UV correction to the three-gluon vertex \eqref{3A_correction_UV}, and $Z_A$ is the wave function renormalization introduced in \eqnref{eq:renormalized_parameters}. 
Note that the wave function renormalization $Z_A$ enters $g_r$ since the vertex correction was computed with the unrenormalized fields $A_\m$, which hence need to be replaced by their renormalized counter parts $A^r_\mu = Z^{-1}_A A_\mu$. 
One eventually obtains $\b\sim-Ng^3<0$, i.e. a $\b$-function with negative sign~\cite{Blaschke:2009e,Martin:1999aq, Armoni:2000xr,Minwalla:1999,Ruiz:2000} which indicates asymptotic freedom and the absence of a Landau ghost.

Finally, the linear IR divergence in the 3-gluon vertex correction, which qualitatively results to $N$ times the expression \eqref{eq:counterterm-3A}, leads to a renormalized parameter
\begin{align}
\g'^2_r&=\g'^2Z_{\g'}Z_A^3
\,, & \text{with } \qquad
1-Z_{\g'}&\propto Ng^2\,.
\end{align}
Hence, it is absorbed by the according counter term.

\section{Conclusion and Outlook}
In this paper, the {\nc} $U_\star(N)$ gauge field action \eqref{eq:renormalizable_action} has been proposed as a generalization of the previously put forward $U_\star(1)$ counter part of Ref.~\cite{Blaschke:2009e}. At least at one-loop level, all dangerous IR divergent terms can be absorbed into according counter terms, and one may hope that this is true also for higher loop orders. Furthermore, the wave-function renormalization $Z_A$ is exactly the same for the $U_\star(1)$ and the $SU_\star(N)$ gauge fields, which is crucial if the present model is to be taken seriously as a candidate for a renormalizable {\nc} $U_\star(N)$ gauge field model. In a next step towards a general proof of renormalizability, however, further explicit (higher-)loop calculations are required.

\subsection*{Acknowledgements}
Discussions with H.~Grosse, H.~Steinacker and M.~Wohlgenannt, as well as correspondence with A.~Armoni are gratefully acknowledged. 
This work was supported by the ``Fonds zur F\"orderung der Wissenschaftlichen Forschung'' (FWF) under contract P21610-N16.


\end{document}